\documentclass[%
reprint,
unsortedaddress,
 amsmath,amssymb,
 aps,
prb,
]{revtex4-2}

\usepackage{graphicx}% Include figure files
\usepackage{dcolumn}% Align table columns on decimal point
\usepackage{bm}% bold math
\usepackage{hyperref}% add hypertext capabilities
\usepackage{xcolor} % colored text to exchange comments.
\begin{document}

\preprint{APS/123-QED}

\title{Single crystal studies and electronic structure investigation of a room temperature semiconductor NaMnAs}
%\thanks{A footnote to the article title}

\author{Ji{}\v r\'i Voln\'y}
\affiliation{Department of Condensed Matter Physics, Faculty of Mathematics
and Physics, Charles University, Ke Karlovu 5, Praha 2, CZ--12116}

\author{V\'aclav Hol\'y}
\affiliation{Department of Condensed Matter Physics, Faculty of Mathematics
and Physics, Charles University, Ke Karlovu 5, Praha 2, CZ--12116}
\affiliation{Institute of Condensed Matter Physics, Faculty of Science,
Masaryk University, Kotl\'{a}\v{r}sk\'{a} 2, Brno, CZ--61137}
\author{Kate\v rina Charv\'atov\'a,
Martin Veis}
\affiliation{Department of Condensed Matter Physics, Faculty of Mathematics
and Physics, Charles University, Ke Karlovu 5, Praha 2, CZ--12116}

\author{M. Vondr\'a\v cek, J. Honolka}
\affiliation{FZU - Institute of Physics of the Czech Academy of 
Sciences, Na Slovance 2, Praha 8, CZ--18221}

\author{Elen Duverger-N\'edellec}
\affiliation{Department of Condensed Matter Physics, Faculty of Mathematics
and Physics, Charles University, Ke Karlovu 5, Praha 2, CZ--12116}
\affiliation{CNRS, ICMCB UMR5026, Bordeaux INP, University of Bordeaux,
Pessac, F--33600}

\author{J. Schusser}
 \altaddress{Experimentelle Physik VII and W\"urzburg-Dresden Cluster of
 Excellence ct.qmat, Universit\"at W\"urzburg, W\"urzburg, D--97074}
 \affiliation{New Technologies-Research Center, University of West Bohemia,
 Plze\v n 3, CZ--30100}
 
%\affiliation{Experimentelle Physik VII and W\"urzburg-Dresden Cluster of
% Excellence ct.qmat, Universit\"at W\"urzburg, W\"urzburg, D--97074}
 
\author{S.W. D'Souza, J. Min\'ar}
\affiliation{New Technologies-Research Center, University of West Bohemia,
 Plze\v n 3, CZ--30100}
 
\author{James M. Pientka}
\affiliation{Department of Physics, St. Bonaventure University, St. Bonaventure, NY--14778} 
 
\author{Alberto Marmodoro, Karel V\'yborn\'y}
\affiliation{Institute of Physics, Academy of Science of the
Czech Republic, Cukrovarnick\'a 10, Praha 6, CZ--16253}

\author{Kl\'ara Uhl\'\i{}\v rov\'a}
\affiliation{Department of Condensed Matter Physics, Faculty of Mathematics
and Physics, Charles University, Ke Karlovu 5, Praha, CZ--12116}

\email{klara@mag.mff.cuni.cz}

\date{Sep07, 2021}

\begin{abstract}
We report synthesis of single crystalline NaMnAs, confirm its
antiferromagnetic order and characterise the sample by photoemission
spectroscopy.
%Physical properties of NaMnAs were studied on single crystalline samples prepared for the first time. 
The electronic structure was studied using optical transmittance, x-ray and ultraviolet spectroscopy and by theoretical modeling using local density approximation (LDA) extended to LDA+U when Heisenberg model parameters were determined. Optical transmittance measurement have confirmed the theoretical predictions that NaMnAs is a semiconductor. Also the N\'eel temperature was closer determined for the first time from temperature dependence of magnetization, in agreement with our Monte Carlo simulations.
\end{abstract}

 %Use showkeys class option if keyword
\maketitle

%\tableofcontents

\section{\label{sec: introduction}Introduction}
Antiferromagnets (AFMs) are increasingly coming into the spotlight for spintronics applications 
\cite{Jungwirth2014,Marti2015,Gomonay2017,Jungfleisch2018}.
One of the motivations is robustness against stray magnetic fields for memory devices applications \cite{Marti2014,Olejnik2017,Wadley2018}, thanks to compensation of magnetic moments.
The aspect of linear spin waves dispersion close to the $\Gamma$ point,
as opposed to the quadratic trend in ferromagnets \cite{Rezende2019}, 
has also been considered for possible magnonics technology \cite{Kruglyak2010a,Grundler2016,Chumak2017,Rezende2019,
Ghader2019,Ross2019,Han2020}.

While numerous AFM materials were discovered in previous century, for many
of them, little is known beyond the bare fact that they are
antiferromagnetic. One of the most widely studied compounds in AFM
spintronics~\cite{Baltz2018,Jungwirth2018}, CuMnAs shows switching 
behavior in charge resistivity between binary \cite{Wadley2016a} 
or multi-level \cite{Olejnik2017} states, through the controlled application
of electric current \cite{Grzybowski2017,Wadley2018a} or optical \cite{Kaspar2019} pulses.
Its tetragonal phase presents the additional benefits of above room temperature 
antiferromagnetic ordering \cite{Maca2012,Wadley2015,Maca2017,Veis2018,Maca2019, Uhlirova2018b},
as well as lack of lattice strain in the deposition over a GaP substrate.

On the other hand, similarly to the other well-established example of Mn$_2$Au,
CuMnAs presents no electronic band gap \cite{Sapozhnik2018}.
This feature would be highly desirable for further spintronics and magnonics
technologies, in which for instance lack of charge carriers have been predicted \cite{Odashima2013} and observed \cite{Kajiwara2010,Cornelissen2015}
to provide long range magnons diffusion length due to the depopulated Stoner continuum.

While the pursuit of better characterization of microscopic features 
and switching mechanisms in the most established AFMs continues,
and typically resorts to high quality samples prepared by molecular beam epitaxy (MBE)
\cite{Jungwirth2011,Wadley2013,Hills2015},
the search for other semiconducting AFMs has also carried on,
taking advantage of the flexibility offered by bulk synthesis methods
in exploring different compositions and lattice geometries.

Remaining within the $A$Mn$X$ family of alkaline metal / manganese-pnictide ternary compounds 
with a tetragonal non-symmorphic space group P4/nmm (Cu$_2$Sb-type structure), we recall the early work by Linowsky and Bronger (with $A$ = K and $X$ = P, As) \cite{Linowsky1974} Schuster et al. (with $A$ = K,Na and $X$ = Sb, Bi, P) \cite{Schuster1978}, Achenbach et al. (with $A$ = Li, Na and $X$ = P, As, Sb, Bi) \cite{Achenbach1981}, Bronger et al. ($X$=Na,Li,K and $X$=P,As,Sb,Bi) \cite{Bronger1986}. The later work showed that all these compound order antiferromagnetically well above the room temperature. These materials have  recently been rediscovered as potential antiferromagnetic semiconductors  \cite{Zhou2016,Beleanu2013,Beleanu2014,Jungwirth2011,Yang2018,Wegner2020}.   
Here, we would like to focus on NaMnAs. Up to now only the crystal structure and magnetic structure by XRD and neutron diffraction, respectively, was reported on pulverised polycrystals [2,6]. It has been shown that it orders antiferomagnetically with magnetic moments aligned along $z-$ axis and propagation vector $k$ = [000]. The N\'eel temperature was estimated between 293 and 643 K; close to zero temperature, the Mn-magnetic moment was determined to be 4.0~$\mu_B$/Mn~\cite{Bronger1986}.

We report on the high quality single crystal growth of NaMnAs, confirm
by optical transmission measurements its semiconducting nature and
discuss various aspects of magnetic order both from experimental and
theoretical point of view. 

\section{\label{sec: sample growth and crystall structure measurement}Sample growth and characterization}

The NaMnAs single crystals were grown similar to Cu$_{1-x}$Mn$_{1+x}$As using the flux method \cite{Uhlirova2015}.  The single crystals were shiny flat rectangular plates ($2\times 2 \times 0.1 \mbox{ mm}^3$), which can be cleaved in the basal plane using a standard sticky tape. This suggests very weak bonds between the layers. A typical sample is shown in Fig.  \ref{fig: NaMnAs after cleaving}. The material is not stable on air. After few minutes it becomes dark so most of the manipulation has to be done under protective atmosphere (see Supplemental Material). It is rather typical for the flux method, that also other phases are grown; in this case MnAs crystals were formed during the synthesis. They form tiny needles which are often attached to the surface of NaMnAs crystals and are difficult to remove. This does not affect any spectroscopic measurements which are performed on impurity-free, cleaved surfaces. They influence however other bulk measurements such as magnetization (as discussed later). 

\subsection{Composition and crystal quality}
The composition and the crystal structure of selected single crystals have been determined by energy-dispersive x-ray spectroscopy (EDS) and by single-crystal x-ray diffraction (XRD), respectively.
EDS was performed using Scanning Electron Microscope (SEM) equipped with an energy dispersive x-ray detector Bruker AXS which utilizes ESPRIT software package (a non-standard method with precision up to 1 \%). The as-grown samples were mounted on a SEM stub using carbon tape, freshly cleaved, and quickly installed into the SEM. The surfaces appear very flat and homogeneous with  the stoichiometry: Na$_{0.96(0.02)}$Mn$_{1.07(0.02)}$As$_{0.97(0.02)}$, reproducible across several growth-batches. 
High resolution XRD using Rigaku Smartlab 45/200 in parallel beam mode was performed on an as grown single crystal with the (00l) planes parallel to the sample holder. The sample was preserved by kapton tape during the measurement. From the refined XRD peaks positions, a Cohen-Wagner plot provides the lattice parameters: $a$ = $b$ = 4.213 $\pm$ 0.002 and $c$ = 7.0955 $\pm$ 0.0005 (\AA).  The higher precision of the $c$-parameter is due to the higher number of measured (00l) diffraction peaks. Also the free z-component of the 2c atomic position was refined being 0.6675 and 0.2166 for Na and As atoms, respectively.  The standard single crystal diffraction collecting the full diffraction pattern was preformed as well, however, the crystals exhibit large degree of mosaicity, preventing proper crystal structure refinement ( See Supplemental Material at ). 

\begin{figure*}[htb]
 \centering
 \includegraphics[width=1.0\textwidth]{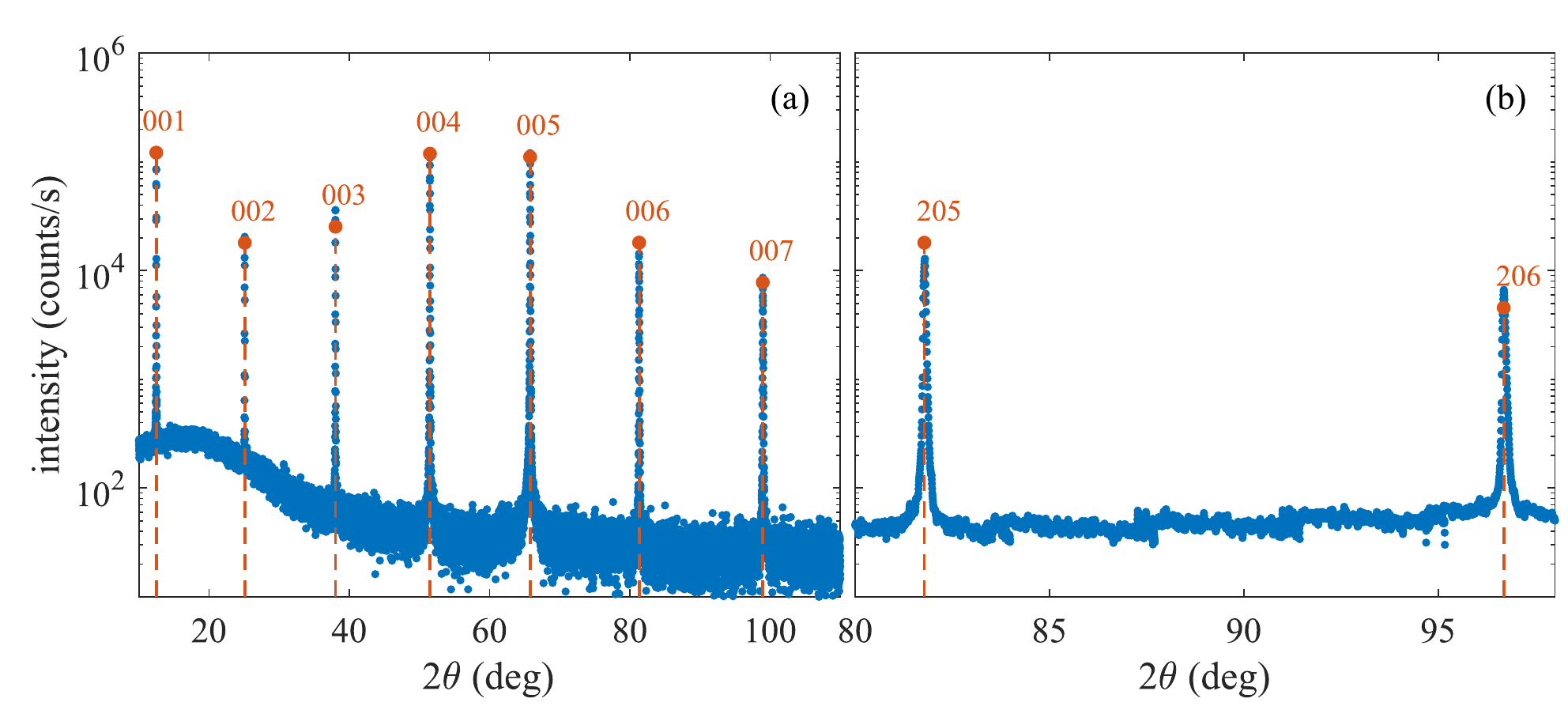}
 \caption{\label{fig: theta scan}
 Symmetric 2$\theta$/$\omega$ scan of the (001) NaMnAs single crystals (a),
 and $Q_z$ scan across the (205) and (206) directions, a and (b), respectively. Open circles represent the calculated diffraction intensities, using the crystal structure from Ref. \cite{Bronger1986}. In the calculation we assumed an ideal kinematically diffracting crystal lattice}
\end{figure*}

\begin{figure}[htp]
 \centering
\includegraphics[width=0.7\columnwidth]{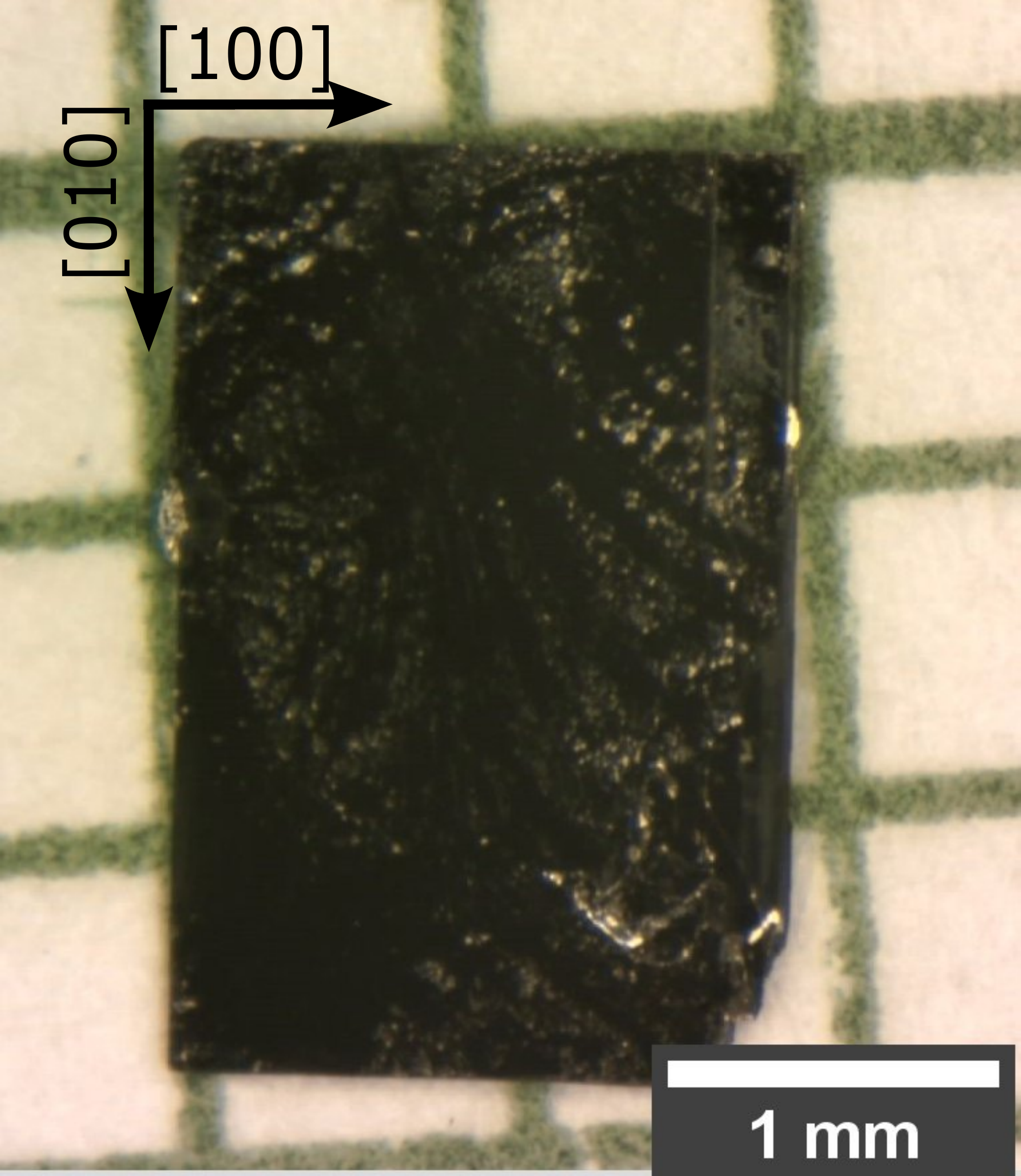}
\caption{\label{fig: NaMnAs after cleaving} Optical microscope image of typical NaMnAs single crystal. The sample thickness in the order of 100 $\mu$m. The $c$-axis is pointing perpendicular to the plane while $a$-axes are parallel with the long edges of the sample.}
\end{figure}

Fig. \ref{fig: theta scan} shows a symmetric 2$\theta$/$\omega$ scan of the (00l) diffraction peaks. The peak width is limited by the resolution of the diffractometer, and it suggests excellent homogeneity of the crystals. Additionally, scans across the (205) and (206) directions were performed in order to determine the $a$ lattice parameter. 

\begin{figure}[htp]
 \centering
 \includegraphics[width=0.7\columnwidth]{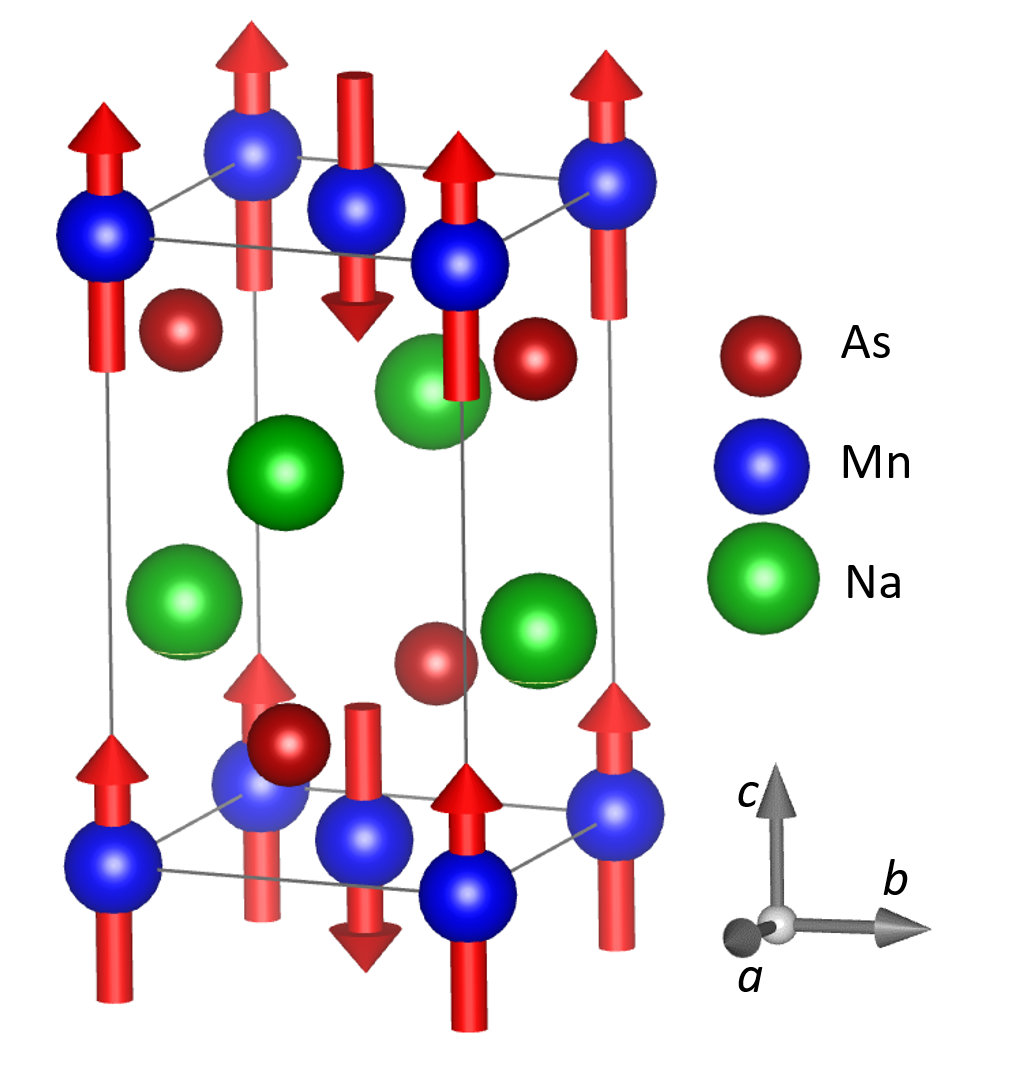}
 \caption{\label{fig: unit cell}Unit cell of NaMnAs with magnetic moment directions refined by Bronger et al. \cite{Bronger1986}.}
\end{figure}

\subsection{X-ray photoemission spectroscopy}

\begin{figure}[htb]
 \centering
  \includegraphics[width=1.0\columnwidth]{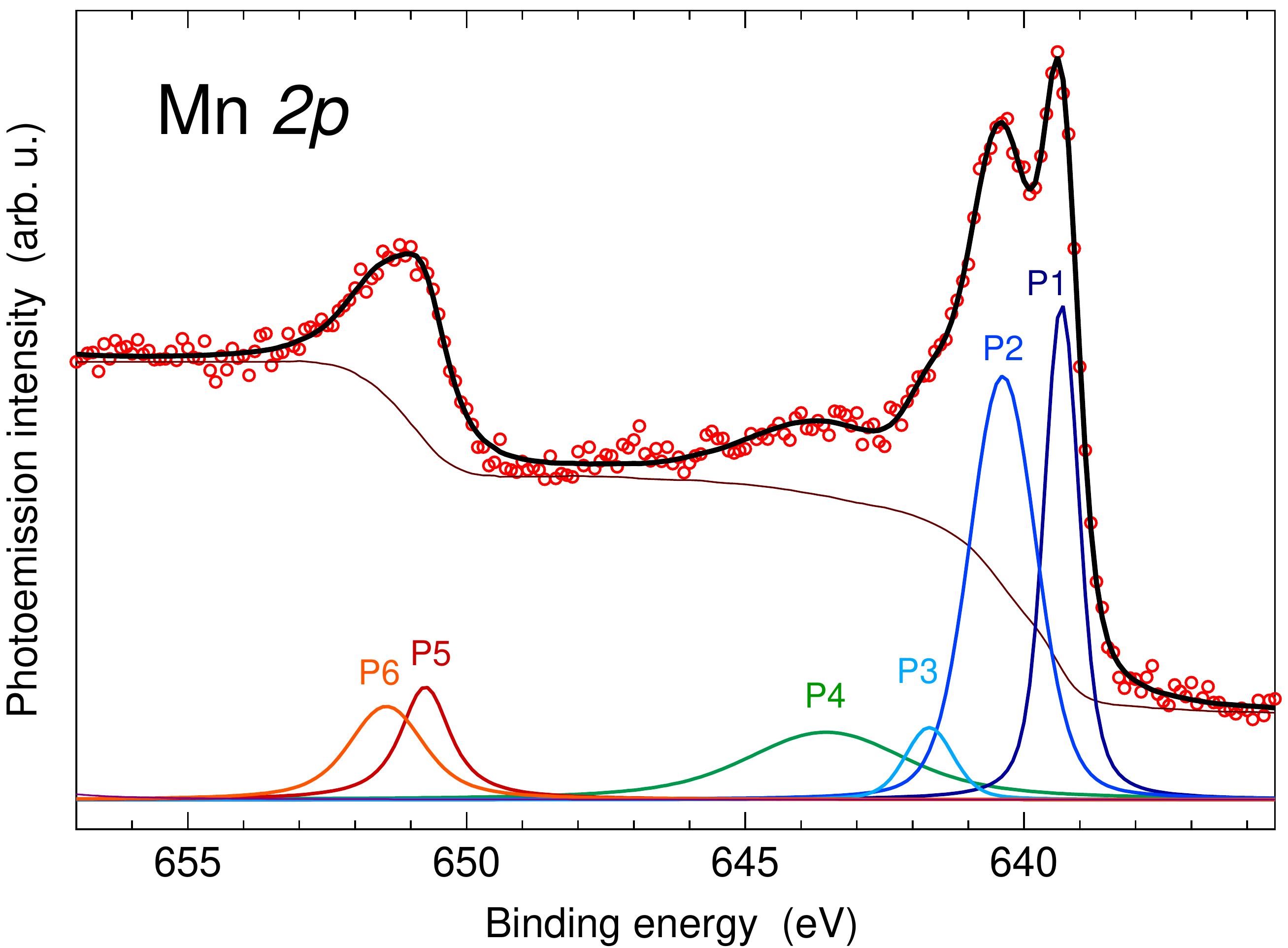}
 \caption{ XPS of the Mn 2$p$ core level. Peaks and the fitting procedure are discussed in 
 the text, the broad peak P4 is the shake-up feature.
}
\label{fig-XPSmain}
\end{figure}

The sample was investigated by x-ray photoemission spectroscopy 
(XPS) after a fresh cleave under UHV conditions.
Sodium, arsenic and manganese shallow core levels (CLs) were probed using
monochromatised Al~K$\alpha$ radiation ($\hbar\omega= 1486.7$~eV), revealing
depth-sensitive information about the sample
composition and its homogeneity as discussed in the Supplemental Material.\\ 
In addition, details in the spectral lineshape of Mn CLs allows an insight into the Mn 3$d$ shell configuration.
The Mn 2$p$ CL in Fig.~\ref{fig-XPSmain} shows a 
complex spectral shape, which can be reproduced using six 
fitted Voigt profiles (referred to as P1--P6) and a ranged Shirley background.
Energy positions, widths, and intensities of the Voigt profiles are summarised
in Tab.~\ref{tbl:XPSvalues}.
The double-peak structure (P1 and P2 separated by 1.1~eV in our case) of the 2$p_{3/2}$ CL is a typical multiplet splitting effect in Mn observed for example in bulk Mn-oxides with Mn$^{2+}$ or Mn$^{3+}$ and even Mn$^{4+}$ states due to strong crystal field effects.\cite{JH-MV-refsMn}
However, such a double-peak structure has also been observed \cite{Tarasenko2015} for single 
Mn atoms embedded in Bi$_2$Te$_3$, where peaks can be reproduced 
in a MnX$_6$ (e.g. X = Br) cluster model\cite{Taguchi} 
with Mn in a 2+ state. Splitting effects in the 2$p_{1/2}$ CL (P5 and P6) are
similar\cite{Tarasenko2015,Taguchi} but broadened due to super-Coster-Kronig decay processes, which limits 
the information on the Mn chemical state. Finally, the smaller spectral feature P3 located at 1.3~eV higher binding energy (BE) compared to P2 may belong to a $J=1,2,3,4$ series of bound states between 2$p_{3/2}$ hole and 3$d$ valence electrons\cite{Wernet2001}. % PRA 63, 050702

\begin{table}[tb!]
\begin{ruledtabular}
   \begin{tabular}{l c c c c c c}
      & $P$1 & $P$2 & $P$3 & $P$4 & $P$5 & $P$6\\
      \hline
          Energy [eV] & 639.33 & 640.38 & 641.70 & 643.46 & 650.73 & 
651.42\\
         FWHM [eV] & 0.75 & 1.41 & 0.97 & 4.05 & 1.06 & 1.60\\
         Area [a.u.] & 169 & 260 & 37 & 114 & 65 & 89\\
    
   \end{tabular}
\end{ruledtabular}
\caption{Peak energy positions, FWHM values, and total intensities of Voigt peaks $P_1$ - $P_6$ in 
Fig.\ref{fig-XPSmain}. For $P_1$ - $P_3$ and $P_5$ - $P_6$ natural linewidths of 
0.3 eV and 0.9 eV were assumed, respectively.}
\label{tbl:XPSvalues}
\end{table}

In the following we will discuss possible charge-transfer (CT) effects, which have been identified e.g. in
MnO spectra\cite{Hariki2017}. Despite the convincingly present feature P4 (we were unable to obtain good fits
when P4 is removed), we believe that a CT peak is absent. This feature found in
Fig.~5 of Ref.~\onlinecite{Hariki2017} is located at significantly higher energies
than P4 in our measurements. On the other hand, shake-up satellite of the
2$p_{3/2}$ CL was reported at lower energies~\cite{Zhao1984}. The absence of the CT 
peak bears witness against ionic character of binding in our samples, consistent 
with ab initio predictions discussed later. We therefore conclude that the object of
our study has grown in the desired composition (without significant amounts of other
phases such as MnAs or manganese oxides) and relegate a more detailed analysis of
the XPS for later investigations.

\subsection{Ultraviolet photoemission spectroscopy}

\begin{figure}[htb]
 \centering
 \includegraphics[width=1.0\columnwidth]{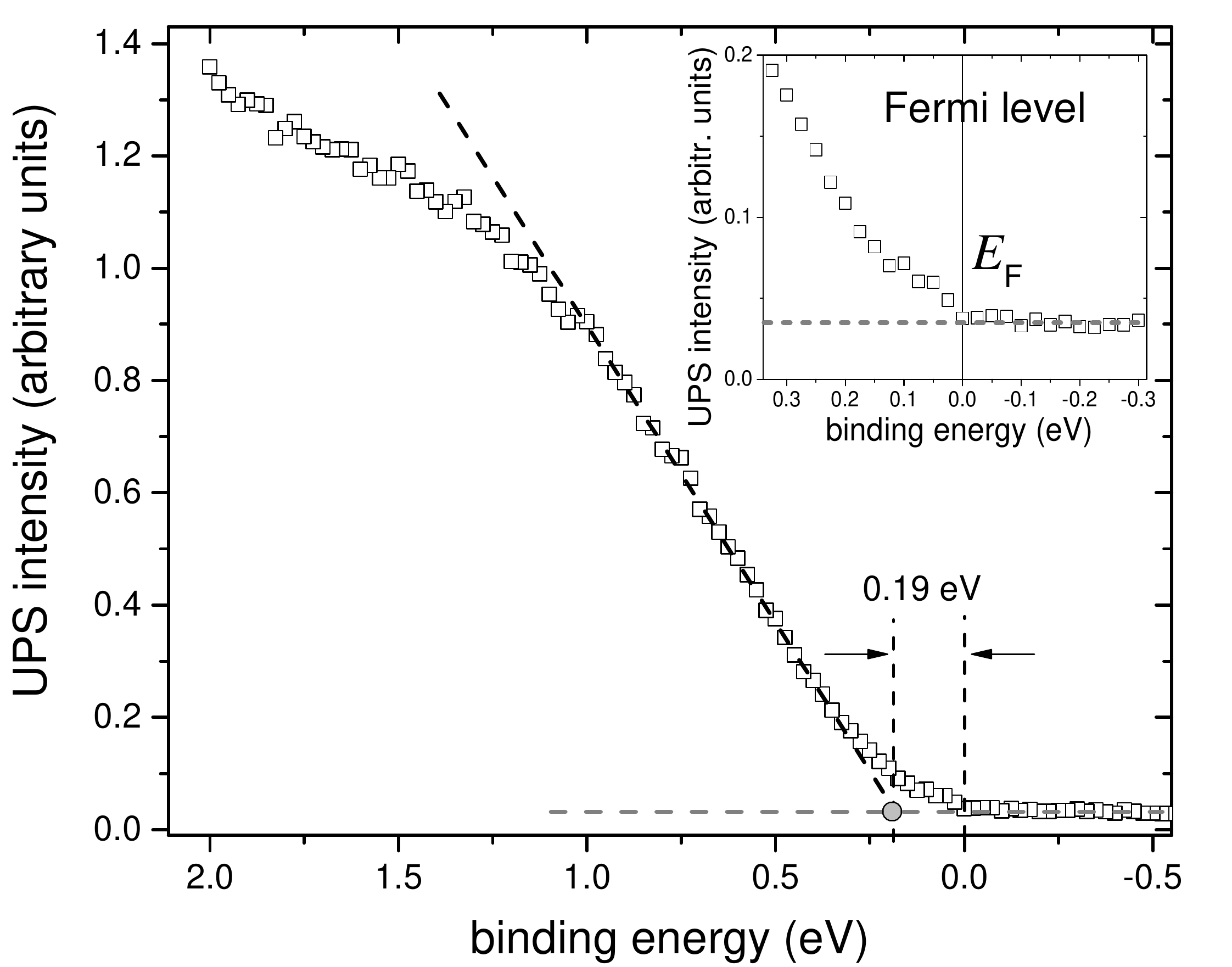}
 \caption{% formerly: \label{fig: UPS}
 UPS data covering the VBM region measured with a photon energy $\hbar\omega$ = 21.2~eV. The position of the VBM at 0.19 eV is estimated by the leading edge method (see text). In the inset the signal intensity around $E =0$ (analyzer Fermi edge) is shown in detail.
}
\label{fig-UPS}
\end{figure}

The same device (as in the case of XPS) with a different photon
excitation energy ($\hbar\omega=21.2$~eV, He~I line) was used to
obtain UPS data shown in Fig.\ref{fig-UPS} covering BEs in the range 
[2~eV, -0.5~eV] measured in normal emission at a photon energy of 21.2~eV. 
The reference point $E =0$~eV corresponds to the Fermi level of the PES analyzer. 
In semiconductors the position of the Fermi energy ($E_{\text F}$) 
with respect to the valence band maximum (VBM) is sensitive to defect states and
 possible surface band bending effects. Kraut {\it et al.}\cite{Kraut1980} described procedures to quantify defect induced shifts in $E_{\text F}$ and band bending in the semiconductor GaAs by referencing shallow CLs such as $E_{\text{Ga 3d
}}$ to the VBM value $E_{\text{VBM}}$. Thereby values $E_{\text{VBM}}$ are estimated using the leading edge method, which approximates the density of states (DOS) by a tangential line at the maximum steepness of the VB edge.\\
In our case the tangential line of the dominant intensity suggests a VBM at $E =
 0.19$~eV. Looking more closely to the UPS intensity at lower BEs (see inset in 
Fig.\ref{fig-UPS}), it is evident that the intensity stretches all the way to $
E = 0$ and then disappears, suggesting a well-defined sample Fermi edge and that
 the Fermi level is intersecting the upper VB states. Therefore, we conclude that our sample
is a p-type semiconductor.
 
\section{Magnetization measurements}
In order to determine the N\'eel temperature $T_{N}$ on NaMnAs single crystals, we have performed high temperature magnetization measurements using VSM Oven option (Quantum Design PPMS). A set of about 15 single crystals, with total mass of 16 mg was placed in the sample holder and fixed by copper foil with magnetic field oriented in the basal plane (see Supplemental Material).  The temperature dependence of magnetization was measured in the range 300-880 K and magnetic field of 1 T.   

\begin{figure}[htb]
\centering
\includegraphics[width=1.0\columnwidth]{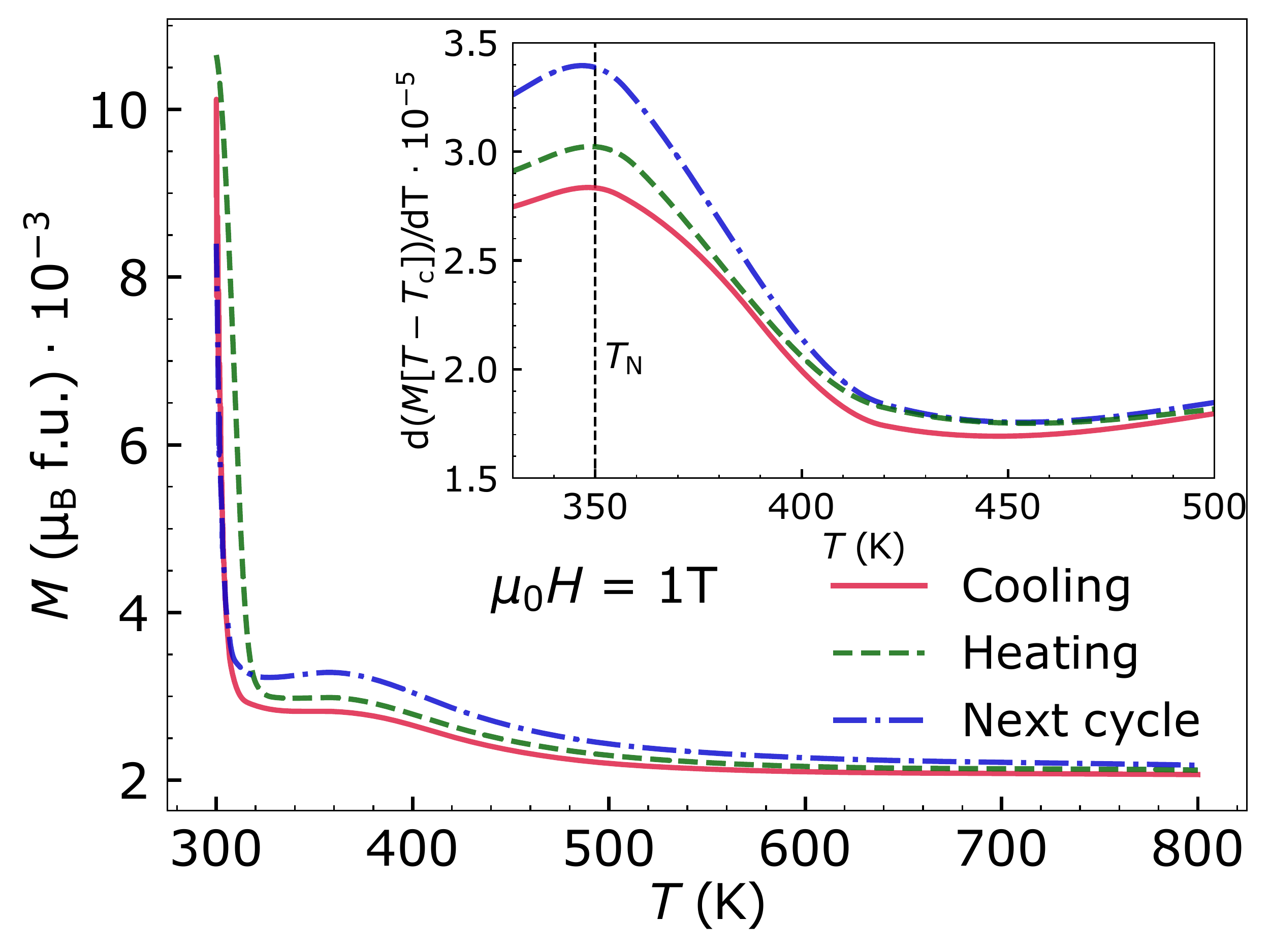}
%{Figures/NaMnAs_magnetizace.png} %moved to dump
\caption{\label{fig: magnetometry} Susceptibility measurements 
(temperature dependent magnetisation at $B=1$~T). The shift of maximum in the
inset, upon thermal cycling, is likely an artefact of the signal-enhancement
method.}
\end{figure}

We have observed, that the overall magnetic signal increases and slowly saturates upon repeating the heating cycles (Fig. \ref{fig: magnetometry}).  This is probably due to the fact, that in vacuum and high temperatures conditions, the sample surface becomes Na deficient and larger amount of MnAs phase is grown, increasing the paramagnetic signal. This hypothesis is supported by XRD measurements of as grown and annealed piece of crystal (see Supplemental Material), where new MnAs peaks appeared after annealing the crystal at similar conditions.  MnAs in its hexagonal form order ferromagnetically at around 315 K with magnetic-filed dependent temperature hysteresis \cite{Koshkidko:2019, Campos2011, Carvalho2010}.

\begin{figure}[htb]
\centering
\includegraphics[width=1.0\columnwidth]{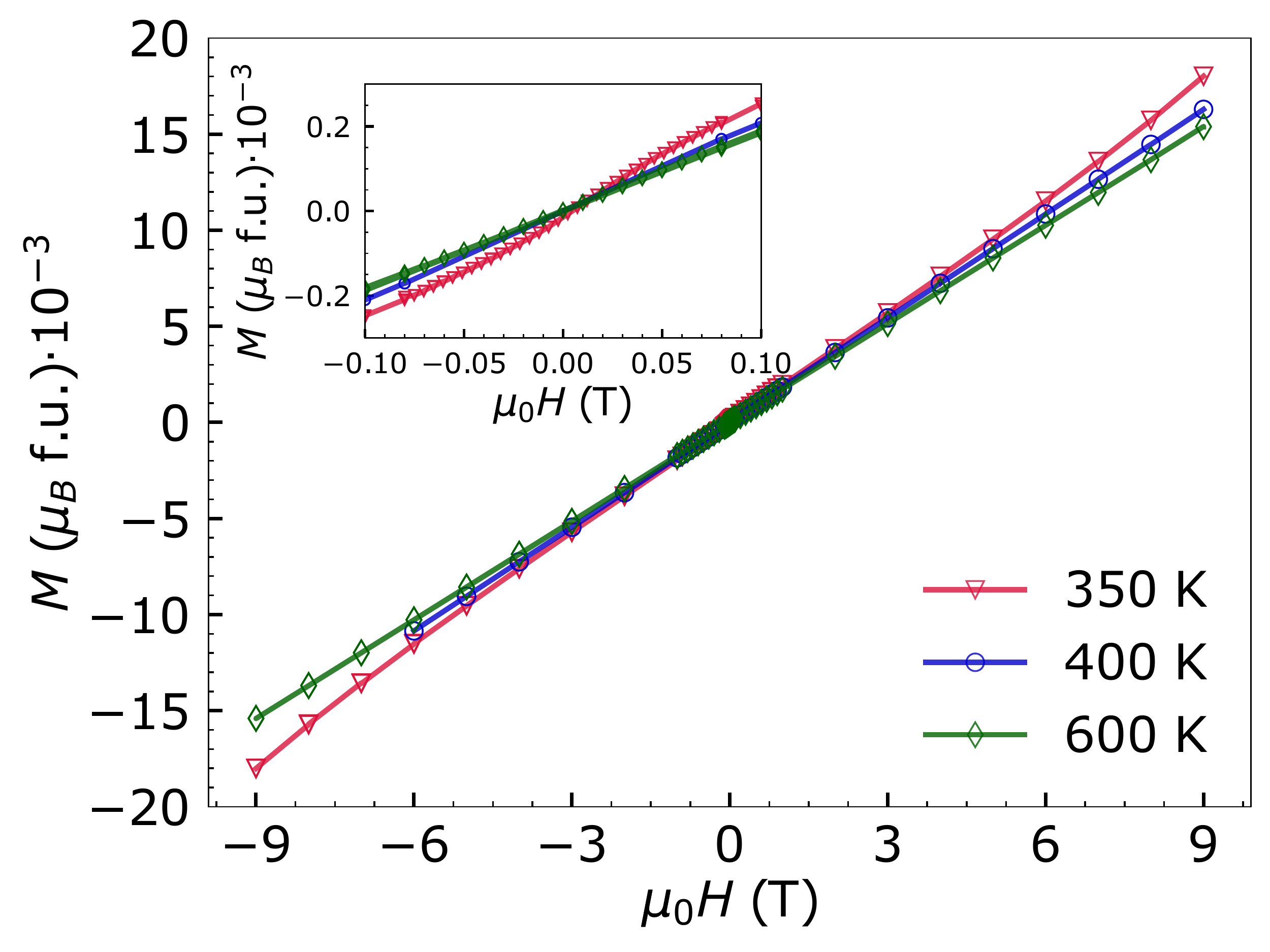}
\caption{\label{fig: magnetization isothemr} Magnetization isotherm of NaMnAs single crystals 
with magnetic field oriented in basal plane.}
\end{figure}

Magnetization isotherm plots (see Fig.~\ref{fig: magnetization isothemr}) at 400, and 600 K show linear behavior without hysteresis, while the the isotherm at 350 K shows a weak paramagnetic contribution and a weak up-turn at high magnetic fields. At lower temperatures (not shown) a hysteresis can be seen connected to the presence of FM impurity of MnAs. 

Considering the presence of MnAs impurities, the magnetization measurement should be considered as partially qualitative. The value of T$_{N}$ fits, however, in the expected range between 295 a 643 K reported by Bronger et al. \cite{Bronger1986}. 

\section{Magnetic order}

Results of the previous Section confirm NaMnAs as an antiferromagnetic
material. Contrary to the simple case of ferromagnets, this however still
allows for many types of magnetic order described by the set of Mn magnetic
moments $\{S_{\mathrm{Mn}}\vec{m}_i\}$ where we assume $|\vec{m}_i|=1$ and
equal magnitude of all Mn moments $S_{\mathrm{Mn}}$. 
From theoretical point of view, the ground state follows from energy
minimisation of
\begin{equation}\label{eq-03}
  E=-\frac12 S_{\mathrm{Mn}}^2\sum_{ij} J_{ij}\vec{m}_i\cdot\vec{m}_j
\end{equation}
once we neglect higher order terms (as discussed in Appendix).
While we relegate a quantitative discussion 
of the Heisenberg parameters $J_{ij}$ to Sec. VI.B, we now qualitatively
describe the implications of this form of energy on the ordering of
classical magnetic moments $\vec{m}_i$ arranged on lattice shown in 
Fig.~\ref{fig: unit cell}. To begin with, let us assume that non-zero $J_{ij}$
occur only for nearest neighbours in the basal plane (intralayer coupling
$J_0$) and along the $c$-axis (interlayer coupling $J_c$). In terms of 
signs, there are then four options, one leading to ferromagnetic order 
(when both $J_0, J_c$ are positive) and the three remaining
options forcing three different types of AFM order. For $J_0>0$ and $J_c<0$,
each Mn layer is ferromagnetically ordered and such AFM structure has never
been identified across $A$Mn$X$ compounds (where $A$ is an alkali metal and
$X$ is As or Sb) as the early neutron scattering measurements by 
Bronger et al.\cite{Bronger1986} revealed. The remaining two options with
$J_0<0$ lead to a checkerboard pattern of moments,
interleaved along the +$z$ or -$z$ out-of-plane direction with propagation
vector either $\vec{k} = (0 0 0)$ or $(0 0 \frac12)$. Bronger et al.
also determined $S_{\mathrm{Mn}}=3.45$ at room temperature in NaMnAs;
it was not clear however, at which temperature the magnetic order collapses
(the only piece of information to this end was that at 643~K, magnetic
signal in neutron scattering data has already vanished).

While for $A$=Li or K the Mn planes are however further coupled
with Heisenberg exchange parameters favoring anti-parallel couplings 
along the (001) axis, effectively resulting in a twice as big
magnetic unit cell, 
in NaMnAs the manganese layers are only weakly coupled to favor
a parallel ordering: in other words, $J_c>0$ should apply this case.
Such simplified qualitative consideration can become inappropriate
when $J_{ij}$ in Eq.~(\ref{eq-03}) do not decay fast enough with the
distance between sites $i$ and $j$ but in case of NaMnAs, as we show
in Sec.~VI.B, the effective magnetic interactions are rather short-ranged.

We conclude this short Section by several remarks on theoretical studies
in this class of materials. Zhou et al. % w13/newMat/NaMnAs/heap-lit/a
\cite{Zhou2016} compared the ground state energy within density 
functional theory (DFT) framework and the energetically most
favourable magnetic order is the one shown in Fig.~\ref{fig: unit cell}. We
confirm this also for DFT+U (as used previously~\cite{Jungwirth2011}
for LiMnAs and KMnAs; see also Sec. VI.A) whereas the energy
cost of doubled magnetic unit cell is at the level of few meV per f.u.

\section{Measurements of the optical gap}

\begin{figure}[htb]
 \centering
 \includegraphics[scale=0.3]{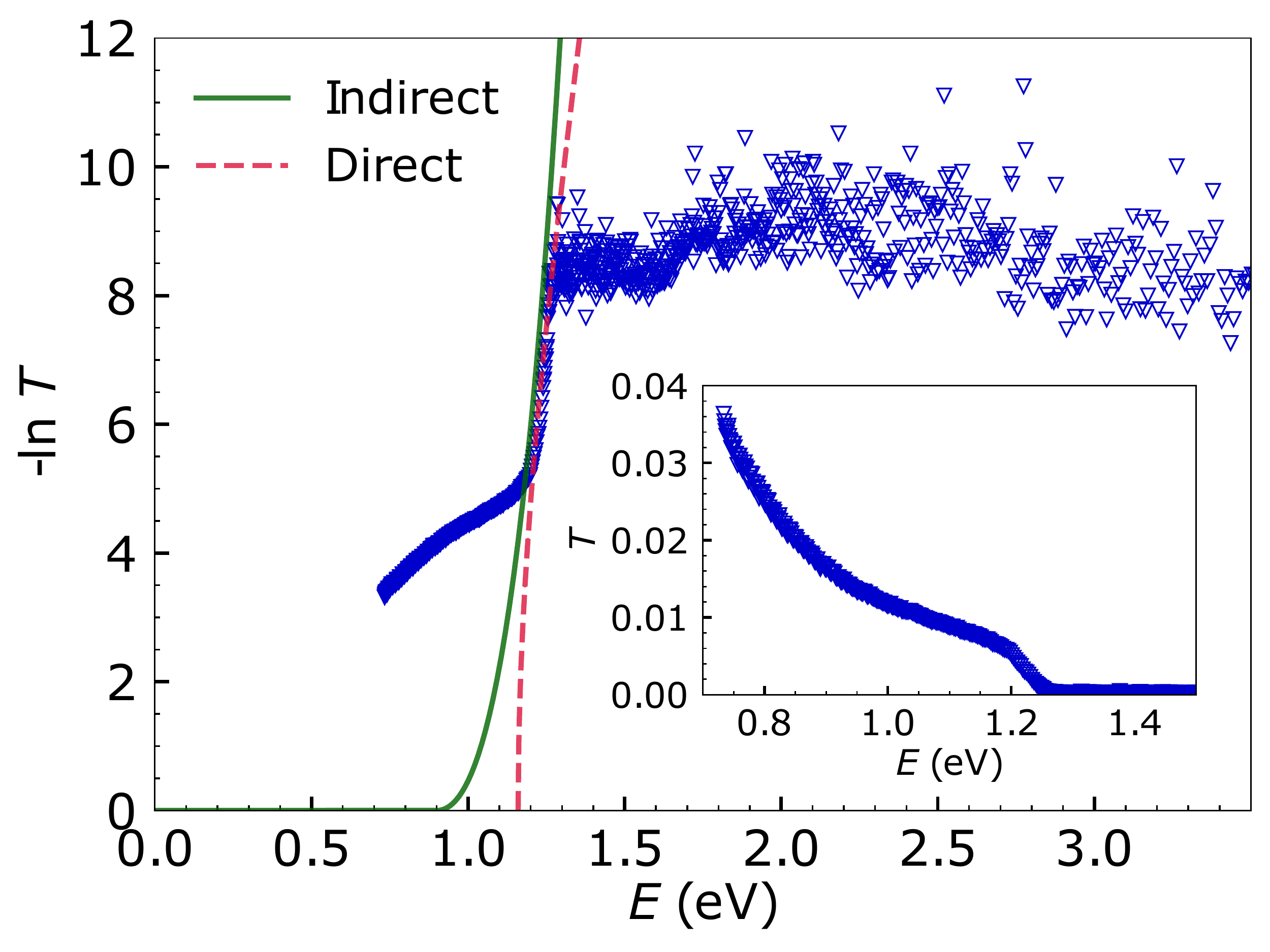}
 \caption{\label{fig: ellipsometry} {\em Inset:} The spectral dependence of the optical transmission $T$. {\em Main plot:} The same data replotted as $-\ln T(E)$; fits in the interval 1.2 - 1.4 eV by direct (red, dashed) or indirect band gap model allow to estimate the band gap.}
\end{figure}

Optical transmission measurements offer a straightforward means to
prove the presence of a band gap.
 The measurements were performed using Woollam RC2 Mueler matrix ellipsometer in the spectral range from 0.7 to 6 eV. The data were corrected for the baseline measurement without the sample in optical path, the resulting transmission data is shown in Fig. \ref{fig: ellipsometry} on a logarithmic scale, with the raw data in the inset. The higher level of noise above 1.2 eV is due to extremely low transmission where the detected light intensity is reaching the sensitivity limit of the equipment. A rapid decrease of transmission towards higher energies near 1 eV is a notable sign of the onset of interband transitions, giving the approximate estimation of the gap. 

Given the definition of $T(E)=I/I_0$, a ratio of incoming and outgoing intensity of light at energy $E=\hbar\omega$, in terms of absorption coefficient $\alpha$, the sample width $w$, the imaginary part of complex permittivity $\epsilon=\epsilon_1+i\epsilon_2$ can easily be extracted from Fig.~\ref{fig: ellipsometry}. We write
$$
- \ln T(E)=w\frac{E}{\hbar c}\cdot \frac{\epsilon_2}{\sqrt{\epsilon_1}}
$$
and fit the experimental data assuming $\epsilon_2\propto\sqrt{E-E_0}$ for a direct and $\epsilon_2\propto (E-E_0)^2$ for an indirect gap. The measurement clearly indicates that there is a band gap: an indirect (direct) band gap fit assuming $\epsilon_1=$const. yields the gap of 0.9 (1.16) eV.
The absorption shoulder at lower energies may be ascribed to defect states inside the gap or twisting of the planes with respect to each other. As already discussed above, although the symmetric XRD scan presented in Fig.~\ref{fig: theta scan} suggest high quality of single crystal, the single crystal diffraction showing the full reciprocal space shows on significant basal plane mosaicity, which can be source of the low-temperature absorption shoulder. Another methods, such as spectroscopic ellispometry can provide wider information about the absorption edge and band structure of the material and will be subject of our further studies of NaMnAs.

\section{Theoretical modelling}

\subsection{Electronic structure}

Magnetic compounds containing manganese in the nominal $3d^5$ configuration
often share a generic electronic structure: while the existence of the gap
is optional, the anion-based bands provide a background for the spin-split
Mn $d$-states one of which is below and the other above Fermi level. 
Previous ab initio calculations\cite{Jungwirth2011,Zhou2016} predicted that
tetragonal XMnAs antiferromagnetic materials (where X is an alkali metal)
do have a gap and, for example, CuMnAs is a metal with low density of
states (DOS) at the Fermi level ($E_F$), see Fig. 3 in Ref.~\cite{Veis2018}.
Size of the gap obtained by ab initio methods can be problematic and density
functional theory (DFT) calculations are known to often give too
small gaps even in simple systems (as opposed to strongly correlated ones) 
such as GaAs.

\begin{figure}
 \includegraphics[width=1.0\columnwidth]{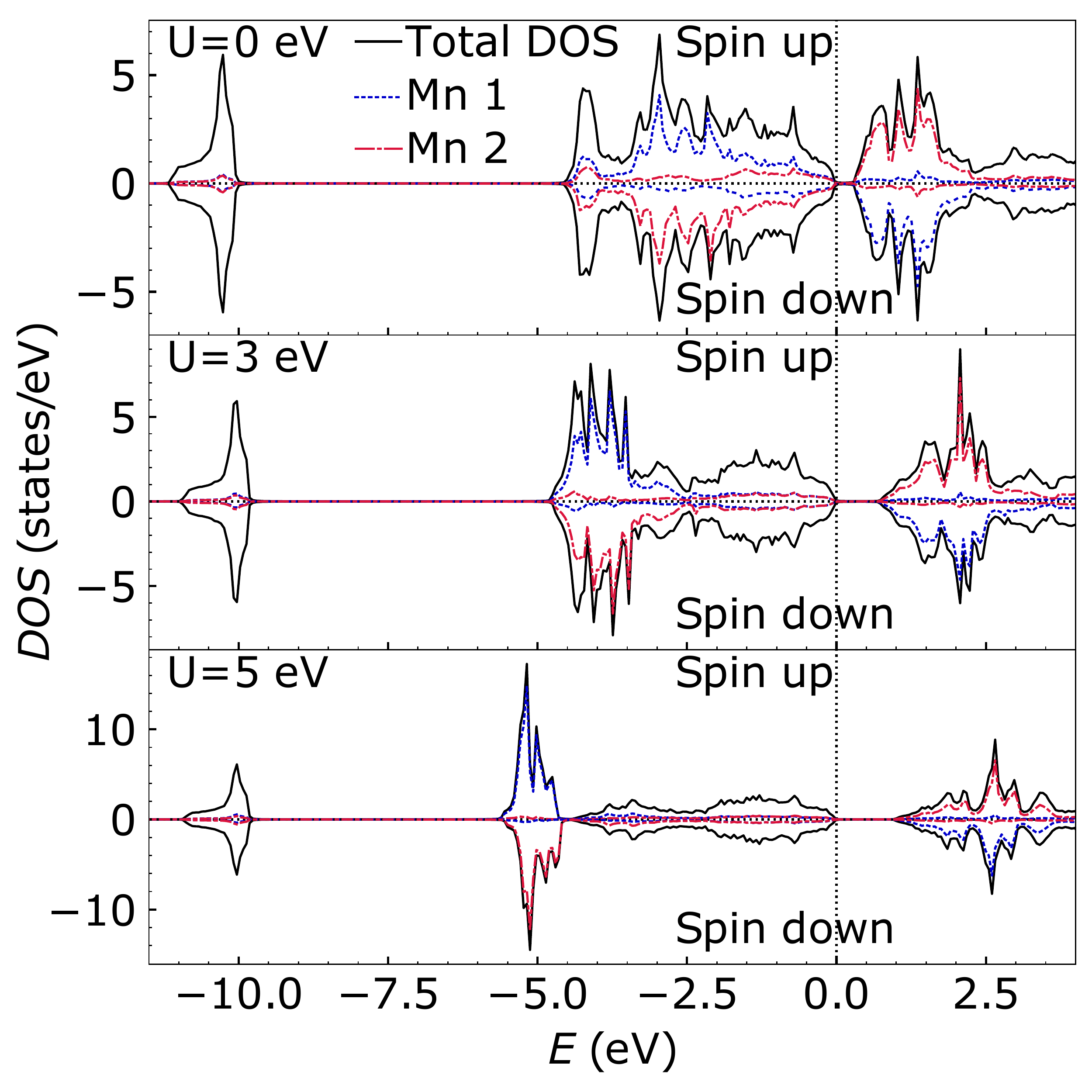}
 \caption{Total and Mn-projected DOS from LDA+U calculations with U = 0, 3, 5 eV.}
 \label{NaMnAs_DOS}
\end{figure}

We chose local density approximations (LDA) as our starting point for DFT.
Apart from an undersized gap, a simple LDA calculation leads to Mn magnetic
moments about 10\% smaller than what is found experimentally. Regardless of the
detailed implementation of DFT+U, magnetic moments become larger when $U$
increases,
%(see Fig.~\ref{fig-elStr}(b))
as a consequence of the larger separation in energy of the majority and
minority Mn $d$-states.
%and inevitable detachment of the Mn states from the valence band. 
Symptoms of such gradual changes can be seen in the density of states (DOS)
%and we point out that when inspecting
shown in Fig.~\ref{NaMnAs_DOS}.
%not only position but also width of the Mn $d$-state peak should be watched. 
A compromise among several factors (magnetic
moments, gap size, estimates of N\'eel temperature discussed in the next
subsection) directed us to a value of $U$ around 5~eV. On the experimental
side, realistic magnitude of Mn magnetic moments~\cite{Bronger1986}
would suggest somewhat smaller $U$ but the final judgement should depend
on a more direct quantity (such as Mn $d$-peak characteristics in UPS).
In the following, we use
%
%seem justified. Our UPS experimental data,
%however, fail to provide any support one way (DFT) or another
%(DFT+U). Also, magnetic anisotropy remains solidly out-of-plane so again, there %is not much reason to go beyond
%DFT unless more advanced ab intio techniques (such as GW)
%are available.
%
%In the following, we use 
both LDA and LDA+U in the atomic sphere approximation (ASA) as implemented
within the spin-polarized relativistic Korringa-Kohn-Rostoker (SPRKKR)
electronic structure package~\cite{Ebert2011} and cross-check the results
using generalised gradient approximation to DFT in linearised
augmented-plane-wave (LAPW) method in another package~\cite{Wien2k}.
The latter yield an out-of-plane magnetic anisotropy
around 0.2 meV per formula unit free of any strong dependence on the value
of $U$ with a modest correction due to dipolar interactions. This is somewhat
larger than in CuMnAs (see App.~A of Ref.~\onlinecite{Volny2020}).
\subsection{Heisenberg parameters}
Effective spin Hamiltonians allow to study finite-temperature properties of
magnetic systems (such as NaMnAs) at a reasonable computational cost (see also
discussion in Appendix). Depending on their complexity, they use various
parameters as input where Heisenberg exchange parameters $J_{ij}$ play 
the central role. In our work, $J_{ij}$ are determined using the 
method of Ref.~\onlinecite{Lichtenstein:1984_a}
i.e. starting from the perturbed  (by $\delta V_i$, see Appendix)
%Eq.~\ref{eq-02})
antiferromagnetic ground state (reflected in the $T$-matrices) 
and evaluating
\begin{equation}
  J_{ij}= \frac1{4\pi} \int_{-\infty}^{E_F} dE\mbox{ Im Tr }
  \delta V_i T_\uparrow^{ij}\delta V_j T_\downarrow^{ji}
  \label{eq-01}
\end{equation}
Herewith, the complicated quantum-mechanical problem (crystal 
with many electrons, for example) is mapped to a relatively simple lattice
model of classical spins with pair interactions only. In this model, total
energy (more precisely, its part --- see Appendix) of the system is approximated by Eq.~(\ref{eq-03}) and using a Monte
Carlo simulation, various thermodynamic quantities can be estimated. For
example, temperature-dependent sublattice magnetisation can be extrapolated 
to zero and thus N\'eel temperature obtained.
%The caveat here is
%that while $J_{ij}$ in Eq.~\ref{eq-01} are determined for the zero-temperature configuration, at high
%temperatures (close to $T_N$), the configuration is very different --- this could also be phrased as, $T=0$ 
%values of $J_{ij}$ may be significantly modified when the system approaches the phase transition.

\begin{figure}[htb]
 \centering
 \includegraphics[width=1.0\columnwidth]{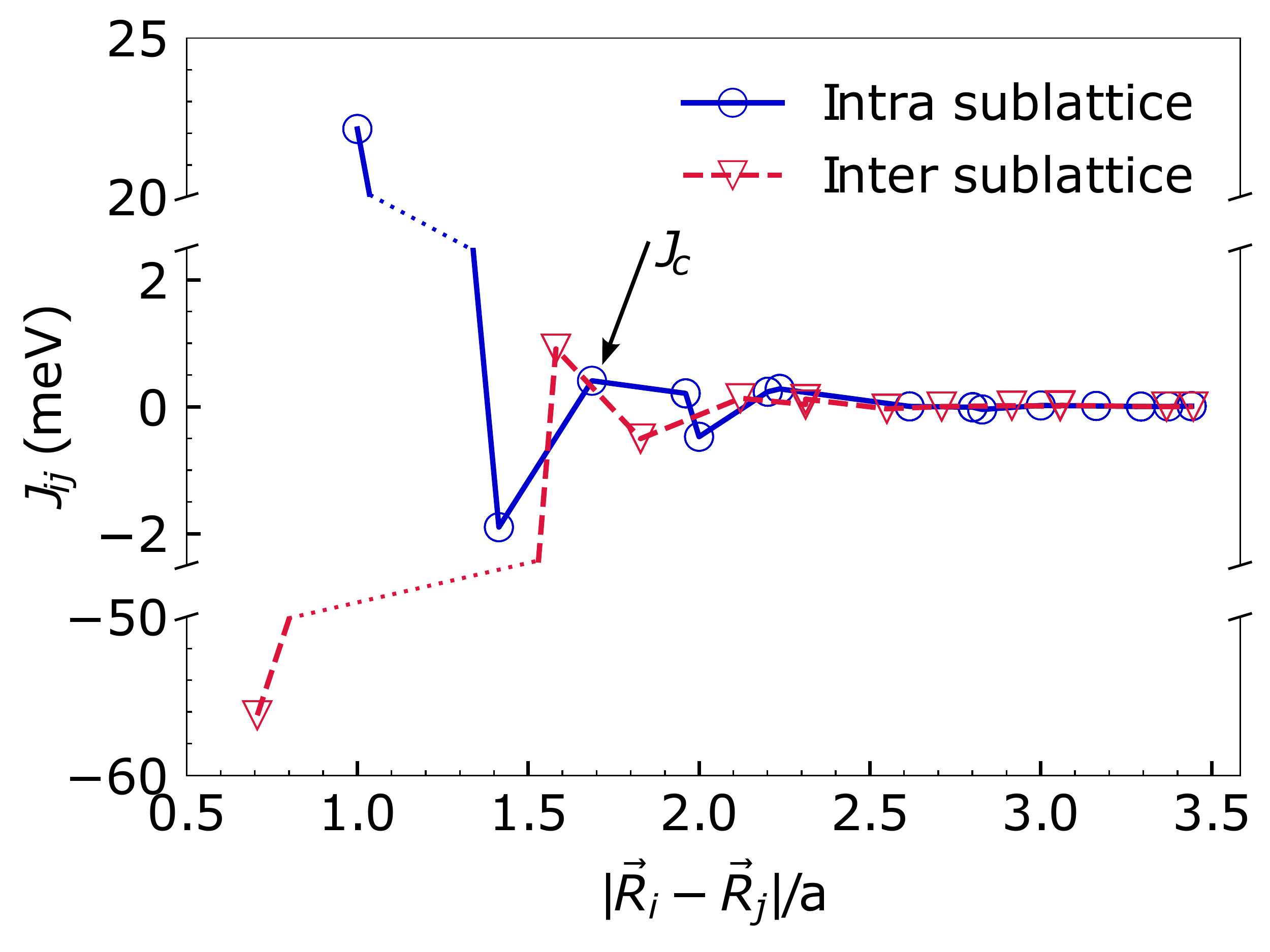}
 \caption{\label{fig: Jij U=5}
% moved to dump/ now and reprocessed it my way (see w13/newMat/NaMnAs/koresp/39)
% \includegraphics[width=8cm]{NaMnAs_JXC_Jij_Mn_1-CUSTOMIZED.eps}
% LRO dataset on: marmodoro@viper:/data/marmodoro/NaMnAs/AF/LDA/Along_z/SPRKKR/NREL-CORE-added_ES/LDA-U_5-AL
Calculated Heisenberg exchange parameters $J_{ij}$ plotted as a function
of distance between sites $i$, $j$.  The arrow indicates $J_{ij}$
corresponding to the leading coupling between layers ($J_c$ in the
qualitative discussion of Sec.~IV).}
\end{figure}

In Fig. \ref{fig: Jij U=5}, we show Heisenberg
parameters obtained from an LDA+U zero-temperature
calculation ($U=5$~eV) plotted as a function of the distance
between individual Mn atoms. This is a simplified
representation of the situation created by an anisotropic
crystal: the complete layout of $J_{ij}$'s in real space is
shown in the Appendix. %Fig.~\ref{fig: Jij LDA+U=5-AL}. 
It should be pointed out that the couplings are clearly dominated by the
nearest-neighbour (NN) pairs within one plane of Mn atoms (both inter- and
intra-sublattice) and the next-nearest-neighbour $J_{ij}$ are more than
an order of magnitude smaller. The antiferromagnetic inter-sublattice NN
coupling (corresponding to $J_0<0$ in terms of the discussion in Sec.~IV)
is further augmented by {\em ferromagnetic} intra-sublattice NN interaction.
More distant pairs interact feebly and coupling between different layers is
also weak. The NN in a different layer, i.e. shortest inter-layer coupling
(corresponding to $J_c$ and highlighted 
by an arrow in Fig.~\ref{fig: Jij U=5}) belongs to
the same magnetic sublattice and $|J_c/J_0|\sim 10^{-2}$.

Comparing LDA to LDA+U calculations, we find that the coupling between Mn 
magnetic moments decreases with increasing $U$. However, this seems 
to be compensated by the enhancement of magnetic moments (see the discussion
of Monte Carlo simulations in the next Subsection).
Zhou et al.\cite{Zhou2016} noted that calculations are robust with respect to 
the alternative choice of GGA rather than LSDA exchange-correlation
functional, thanks to the over-binding artefact of the latter cancelling out in taking the difference.
The reduced dimensionality of the magnetic interactions, mainly confined to the (001) Mn planes, can explain the much smaller odering temperature revealed by experiments and by Monte Carlo simulation, as opposed to cruder early estimates in terms of total energy difference alone, or mean-field theory calculations.

\begin{figure}[htb]
 \centering
 \includegraphics[width=1.0\columnwidth]{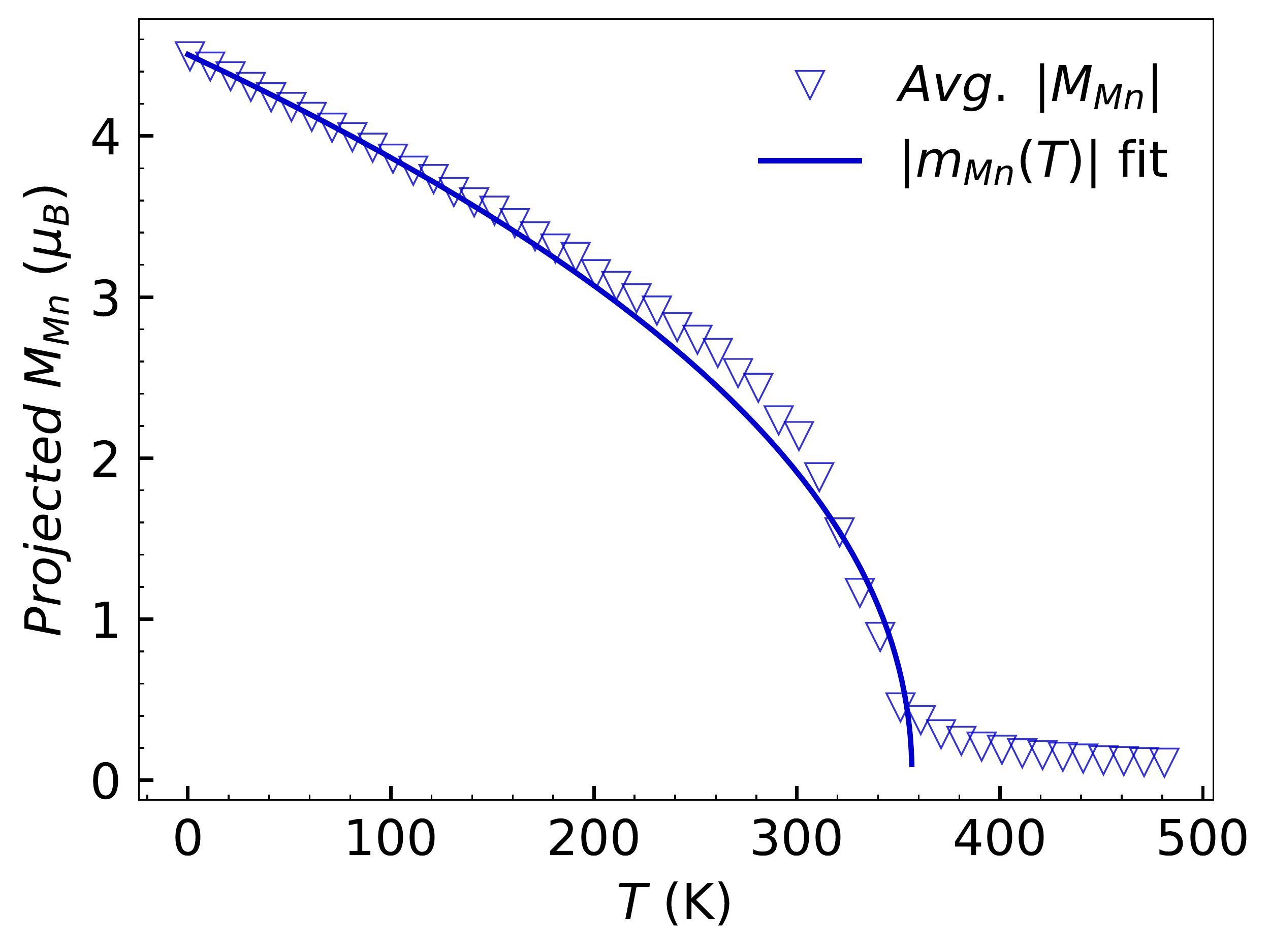}
 \caption{\label{fig: M(T) LRO U=5}Mn-resolved average spin magnetic moment (blue triangles) vs. temperature, from Monte Carlo simulations (see text).
 %(see Tab.\ref{tab: DOS gap vs U}).
 % dataset on: marmodoro@viper:/data/marmodoro/NaMnAs/AF/LDA/Along_z/SPRKKR/NREL-CORE-added_ES/MC_longer_and_larger/LDA-U_5-AL
 }
\end{figure}

\subsection{N\'eel temperature}

%Mermin--Wagner theorem puts limits on the existence of magnetic order
%in two-dimensional systems~\cite{Halperin:2019_a}; it is worth asking 
%which assumptions of this theorem are violated for NaMnAs 
%so that AFM ground state survives as far as beyond the room temperature. %Given that the coupling between Mn layers in NaMnAs is weak but non-zero,
%one can argue that the system is not purely 2D.
%%%%Earlier (see Fig.~\ref{fig-elStr}), we have shown 
%In Sec.~VI.~A we reported that NaMnAs is endowed with a pronounced 
%out-of-plane anisotropy and also the question of interaction range needs
%to be addressed. Below, we first describe our most advanced approach
%to determine the ordering temperature and then proceed to analyse the
%key factors in some detail.

Several approaches to estimating theoretically the ordering
temperature are possible and their short review is
given in Appendix. Here, we employ
Monte Carlo (MC) calculations\cite{Eriksson2017}
as implemented the Uppsala Atomistic Spin Dynamics 
\textrm{UppASD} code \cite{Hellsvik2019}. 
Their input are the Heisenberg
parameters discussed in the previous subsection. 

The lower $T_N$ of NaMnAs with respect to CuMnAs can be tentatively understood as a consequence 
of different effective coordination, or local environment,
for the Mn atoms within the two materials.
While for manganese pnictide CuMnAs the transition metal (Mn atoms) is arranged in a rumpled geometry (2c positions in the CuSb$_2$-type structure),
in NaMnAs the manganese creates flat planes (2a positions of the CuSb$_2$ type stucture).
In combination with the fairly short-ranged $J_{ij}$s
which present only weak coupling across such planes,
the magnetic Hamiltonian becomes effectively 2D.

Temperature-dependent sublattice magnetisation for NaMnAs shown in 
Fig.~\ref{fig: M(T) LRO U=5} suggests that ordering temperature is
slightly above 350~K. This agrees well with experimental data in 
Fig.~\ref{fig: magnetometry}. % and since truncation of Heisenberg
%parameters to nearest neighbours only does not lead to any dramatic
%decay of $T_N$ (not shown), we are left with two options of which
%violation of the Mermin-Wagner theorem assumptions is the crucial one:
%anisotropy and 3D character.

\section{Conclusions}
We have prepared the room temperature antiferromagnet NaMnAs for the first time in the single crystalline form. The single crystals have layered tetragonal structure and can been easily cleaved within the basal plane, the single crystals are in general very soft. The magnetization measurement showed the N\'eel temperature to be 350 K being in agreement with the temperature ranged defined by previous neutron diffraction experiments on polycrystals \cite{Bronger1986}. Using Ultraviolet photoemission spectroscopy and optical transmittivity measurements we showed that NaMnAs is a p-type semiconductor with a band gap between 0.9 and 1.16 eV depending on whether the band gap is direct or indirect, respectively. The detail studies of the type of band gap and band structure in general will be subject of our further work. 
In conclusion NaMnAs has been found to be a new room temperature antiferromagnetic semiconductor, together with the layered structure it can bee promising material for fabrication of functional devices. However, further studies of basic physical properties and possibilities to modify its electronic properties e.g. by chemical substitutions of formation of multi layers has to be done prior to looking into application sphere. 

\begin{appendix}

\section*{Acknowledgements}
We gratefully acknowledge computational resources 
from the Information Technology for Innovation (IT4I) grants: 
OPEN-19-45, OPEN-20-12
and from the project "e-Infrastruktura CZ" (e-INFRA LM2018140) provided within the program Projects of Large Research, Development and Innovations Infrastructures, Operational Program Research, Development and Education financed by European Structural and Investment Funds and the Czech Ministry of Education, Youth and Sports (Project MATFUN - CZ.02.1.01/0.0/0.0/15\_003/0000487). 
Single crystals growth and characterization was performed in MGML (mgml.eu), which is supported within the program of Czech Research Infrastructures (project no. LM2018096).
M.V. and J.H. acknowledge support from
the Czech Ministry of Education, Youth and Sports (Project No.
SOLID21-CZ.02.1.01/0.0/0.0/16\_{}019/0000760). J.M. acknowledges support from GA\v CR Project No. 2018725S.

\end{appendix}

%\setcitestyle{super,open={},close={}} 
\bibliography{Manuscript-Sept07}

% Produces the bibliography via BibTeX.
%\bibliography{references}

\end{document}